\documentclass[12pt]{article}
\begin{document}
\baselineskip=16pt

\title
{\begin{Huge}
Weighing The Black Holes Of GW150914\\
\end{Huge}
\vspace{0.3in}  }
\author{Yuan K. Ha\\ Department of Physics, Temple University\\
 Philadelphia, Pennsylvania 19122 U.S.A. \\
 yuanha@temple.edu \\    \vspace{.1in}  }
\date{May 15, 2017}
\maketitle
\vspace{.1in}
%\begin{center}
%\begin{large}
%Abstract
%\end{large}
%\end{center}
\begin{abstract}
\noindent
We evaluate the mass of the black holes of GW150914 at their event horizons via quasi-local energy approach and obtain
the values of 71 and 57 solar masses, compared to their asymptotic values of 36 and 29 units, respectively, as 
reported by LIGO. A higher mass at the event horizon is compulsory in order to overcome the huge negative gravitational
potential energy surrounding the black holes and allow for the emission of gravitational waves during merging.
We estimate the initial mass of the stars which collapsed to form the black holes from the horizon mass and obtain the impressive values of 95 and 76 solar masses for these progenitor stars.\\

\vspace{0.5in}
\noindent
Honorable Mention Award - Gravity Research Foundation 2017\\
\end{abstract}

\newpage

The discovery of gravitational waves GW150914 by LIGO confirmed the existence of black holes [1]. Black holes are now real astrophysical bodies whereas previously they belonged to the realm of science fiction and physicists' imagination.
They may be abundant in the Universe and their properties can be investigated from the gravitational waves emitted in binary black hole merging.\\

The merging of two black holes and the release of gravitational waves is by nature a process occurring near the horizon.
It is evident that the energy of gravitational waves originates from outside the horizon and not from the interior of the black hole. The signals from GW150914 confirm this is the case. In order for gravitational waves to reach infinity from the vicinity of the horizon, the waves must be significantly red-shifted depending on how close they were generated from the horizon. The waves had a much higher energy when they were emitted than when they are detected. The mass of the black hole at the event horizon is crucial for accounting the redshift of the escaping waves. 
A higher mass at the horizon is compulsory in order to overcome the huge negative gravitational potential energy surrounding the horizon so that a distant observer may detect a black hole with a positive mass. 
Without a higher mass at the event horizon and its neighborhood, there can be no gravitational waves emitted in black hole merging.\\ 

For this investigation, we conclude that the mass at the event horizon of the black holes of GW150914 is significantly higher than the reported values of 36 and 29 solar masses inferred from orbiting motion. We calculate that the masses of the black holes at their event horizons are actually 71 and 57 units respectively. These masses would become the mass of the cores of the collapsing stars which formed the black holes and in turn provide an estimate of the mass of the initial stars without exhaustive numerical simulations. We describe how these mass values are obtained in the quasi-local energy approach [2].\\

The basis for this calculation is the Horizon Mass Theorem for black holes established in 2005 [3]. The theorem is the 
final outcome of the quasi-local energy approach applied to black holes. The quasi-local energy concept is the
most important development in general relativity in the last 25 years to understand the dynamics of the gravitational field [4]. The quasi-local energy is given in terms of the total mean curvature of a surface bounding a volume for a gravitational 
system in four-dimensional spacetime. It agrees with the canonical energy of Arnowitt-Deser-Misner approach [5] at spatial infinity and is naturally suited for investigating the energy distribution of a black hole. The quasi-local energy avoids the ambiguity of defining consistently the energy-momentum tensor locally in general relativity, a problem which has confounded the subject since its beginning.\\ 

The Brown and York expression for quasi-local energy is given in the form of an integral [2]
\begin{equation}
E = \frac{c^{4}}{8\pi G} \int_{^{2}B} d^{2}x \sqrt{\sigma} (k-k^{0}) ,
\end{equation}
where $\sigma$ is the determinant of the metric defined on the two-dimensional surface $^{2}B$ ;
$k$ is the trace of extrinsic curvature of the surface and $k^{0}$ , the trace of curvature of a
reference space. For asymptotically flat reference spacetime, $k^{0}$ is zero.\\

The mass of a black hole depends on where the observer is. The closer the observer gets to the black hole the less gravitational potential energy the observer will see. Since gravitational potential energy is always negative, 
the mass of a black hole increases as one gets near the horizon.
The Horizon Mass Theorem can be stated in the following:\\

\noindent
{\bf Theorem:} 
{\em For all black holes; neutral, charged or rotating, the horizon mass is always twice the irreducible mass observed at infinity}.\\

\noindent
In notation, it is simply
\begin{equation}
M_{horizon} = 2M_{irr} ,
\end{equation}
where $M_{irr}$ is the irreducible mass of the black hole. The theorem is derived only with the knowledge of the spacetime metrics of Schwarzschild, Reissner-Nordstr\"{o}m and Kerr without any further assumption. 
It is an exact and remarkable result in general relativity.\\

In order to understand the Horizon Mass Theorem, it is useful to introduce the following definitions of mass:

\begin{enumerate}
  \item The {\em asymptotic mass} is the mass of a neutral, charged or rotating black hole
        including electrostatic and rotational energy. It is the mass observed at infinity
        used in the various spacetime metrics.
  \item The {\em horizon mass} is the mass which cannot escape from the horizon of a neutral,
        charged or rotating black hole. It is the mass observed at the horizon.
  \item The {\em irreducible mass} is the final mass of a charged or rotating black hole when
        its charge or angular momentum is removed by adding external particles to the 
        black hole. It is the mass observed at infinity.
\end{enumerate}

For a Schwarzschild black hole, the quasi-local energy expression in Eq.(1) gives the total energy contained in a sphere enclosing the black hole at a coordinate distance $r$ as [2,3,6]\\
\begin{equation}
E(r) = \frac{rc^{4}}{G} \left[1 - \sqrt{1-\frac{2GM}{rc^{2}} }\right],
\end{equation}
where $M$ is the mass of the black hole observed at infinity, $c$ is the speed of light and $G$ is
the gravitational constant. At the horizon, the Schwarzschild radius is $r = R_{S} = 2GM/c^{2}$ .
Evaluating the expression in Eq.(3), we find that the metric coefficient 
$g_{00}=(1-2GM/rc^{2})^{1/2}$ vanishes identically and the energy at the horizon is therefore
\begin{equation}
E(r)=\left( \frac{2GM}{c^{2}} \right) \frac{c^{4}}{G} = 2Mc^{2} .
\end{equation}
The horizon mass of the Schwarzschild black hole is simply twice the asymptotic mass
$M$ observed at infinity. The negative gravitational energy outside the black hole has a magnitude as
great as the asymptotic mass.\\

We may consider the total energy of a Schwarzschild black hole contained within a radius at coordinate $r$ to consist
of two parts: the horizon energy which is positive and the gravitational energy which is negative. 
From Eq.(3), it can be evaluated that the total energy at $r = 2 R_{S}$ is $1.17 Mc^{2}$ and at $r = 3 R_{S}$ it is 
$1.10 Mc^{2}$. Thus $ 90\% $ of the gravitational potential energy lies within a distance of 2 Schwarzschild radii outside the horizon and it is within this region that most of the gravitational waves are emitted from black hole merging. At a distance of $r = 10 R_{S}$ , the total energy is only $ 2.6 \% $ higher than the asymptotic value  $ Mc^{2} $ . At $r = 100 R_{S}$, the total energy is  $ {0.25 \%}$ higher than the asymptotic value and one is approaching flat spacetime. 
For an observer outside a solar mass black hole with a Schwarzschild radius of 3 km , nearly flat spacetime is about 300 km away. The two black holes of GW150914 were initially at $ 6.4 \times 10^{7} $ km apart compared to their Schwarzschild radii of approximately 100 km. The two black holes were orbiting each other in flat spacetime with miniscule gravitational wave energy being emitted.\\

For a rotating black hole, we divide the problem into slow rotations and fast rotations. The Horizon Mass Theorem
applies to all cases. The total energy of a slowly rotating black hole with angular momentum $J$ and angular
momentum parameter $\alpha = J/Mc$ in the quasi-local energy approach is given by the
approximate expression [7], 
\begin{eqnarray}
E(r) & = & \frac{rc^{4}}{G} \left[ 1 - \sqrt{ 1 - \frac{2GM}{rc^{2}} + \frac{\alpha^{2}}{r^{2}} } \right ] \nonumber \\
     &   & + \frac{\alpha^{2}c^{4}}{6rG} \left [ 2 + \frac{2GM}{rc^{2}} 
        + \left ( 1 + \frac{2GM}{rc^{2}} \right ) \sqrt{ 1 - \frac{2GM}{rc^{2}} + \frac{\alpha^{2}}{r^{2}} } \right ] + \cdots    \end{eqnarray}
Using the horizon radius for a Kerr black hole,
\begin{equation}
r_{h} = \frac{GM}{c^{2}} + \sqrt { \frac{G^{2}M^{2}}{c^{4}} - \frac{J^{2}}{M^{2}c^{2}} }
\end{equation}
and the definition of the irreducible mass
\begin{equation}
M_{irr}^{2} = \frac{M^{2}}{2} + \frac{M^{2}}{2} \sqrt { 1 - \frac{J^{2}c^{2}}{G^{2}M^{4}} } ,
\end{equation}
we find a very good agreement with twice the irreducible mass as the horizon energy
\begin{equation}
M_{h} \simeq 2 M_{irr} + O(\alpha^{2}) . 
\end{equation} 
Equation (5) is useful for calculating the total energy of a slowly rotating black hole at a coordinate $r$.\\

For general and fast rotations, the quasi-local energy approach has limitation but the total
energy can be obtained very accurately by numerical evaluation in the teleparallel 
formulation of general relativity [8]. The teleparallel gravity is an equivalent geometric formulation
of general relativity in which the action is constructed purely with torsion without curvature. 
It has a gauge field approach. There is a perfectly well-defined
gravitational energy density and the result agrees very well with Eq.(2) at any rotation.
The small discrepancy is due to the axial symmetry of a rotating black hole compared
with the exact spherical symmetry of a Schwarzschild black hole. The result shows that
the rotational energy appears to reside almost completely outside the black hole.\\

For an exact relationship, however, we have to employ a formula known for the area of
a rotating black hole valid for all rotations in the Kerr metric [9],
\begin{equation}
A = 4 \pi ( r_{h}^{2} + \alpha^{2} ) = \frac {16 \pi G^{2}M_{irr}^{2}}{c^{4}} .
\end{equation}
This area is exactly the same as that of a Schwarzschild black hole with asymptotic mass
$M_{irr}$ . Now a local observer who is comoving with the rotating black hole at the event
horizon will see only this Schwarzschild black hole. Since the horizon mass of the
Schwarzschild black hole is $2M_{irr}$ , therefore the horizon mass of the rotating black hole
is exactly $M_{h} = 2M_{irr}$ .\\

For the black holes of GW150914, the primary black hole has a mass of $ 36 M_{Sun} $  with average spin parameter $ a = 0.32 $ . The secondary black hole has a mass of $ 29 M_{Sun} $  with average spin parameter $ a= 0.44 $. They merged to form a final black hole of mass $ 62 M_{Sun} $  and a spin parameter $ a = 0.67 $ . Accordingly, $3 M_{Sun}$  of energy is released as gravitational waves to infinity, i.e.
\begin{equation}
36 M_{Sun} + 29 M_{Sun} = 62 M_{Sun} + 3 M_{Sun}  .
\end{equation}
As remarked earlier, the $3 M_{Sun}$  energy has been significantly red shifted. It is not the initial energy of the waves at their source. According to a previous black hole energy theorem [10], the energy required to remove one unit of mass or inertia $m$ from near the horizon to infinity is exactly the same amount of mass in energy $ mc^{2} $ . Wave energy {\em is} inertia. Thus the redshift of the gravitational waves emitted from near the horizon has the same magnitude as the wave energy observed at infinity, i.e. 
an amount of  $ 3 M_{Sun} $ of energy is expended when $ 3 M_{Sun} $ of wave energy is observed at infinity. The total wave energy at the source is therefore $ 6 M_{Sun} $.\\ 

To find the mass of the black holes at the event horizon, we use Eq.(7) in the form
\begin{equation}
2 M_{irr} =  \left[ 2M^{2} + 2M^{2} \sqrt{ 1 - a^{2} } \right]^{1/2}.
\end{equation}
For the primary black hole, the horizon mass is found to be $71 M_{Sun}$ . For the secondary black hole, the horizon mass is 
$ 57 M_{Sun}$ . The final black hole has a horizon mass of $ 116 M_{Sun}$ .
In an ideal merging, the horizon mass would follow the energy equation
\begin{equation}
71 M_{Sun} + 57 M_{Sun} = 116 M_{Sun} + 12 M_{Sun} .
\end{equation}
Of the $ 12 M_{Sun}$ in this equation,  $ 6 M_{Sun}$ are for gravitational waves and their redshifts; another $ 6 M_{Sun} $ 
are for uncertainties. The quasi-local energy approach is an important new tool in understanding black holes dynamics. Without a higher mass at the event horizon, there can be no gravitational wave emission in black hole merging. The asymptotic mass does not account for the redshift; it is an effective mass indispensable for analyzing orbits at large distances.\\

We finally point out that the horizon mass can provide an analytic way of determining the mass of the core of the stars if they collapsed to form the black holes of GW150914. The core of the original stars must have at least the horizon values of $71 M_{Sun}$  and  $57 M_{Sun}$ . This result puts the initial mass of the progenitor stars immediately above $40 M_{Sun}$ , a value which is accepted as the minimum mass of a star which would end up as a black hole. The masses of the progenitor stars are much higher. Giant stars have large cores. The core mass of a primeval star with low metallicity is high due to low mass loss. If the core mass is $3/4$ of the total star mass, then this would lead to an impressive value of $95 M_{Sun}$  for the bigger star and  $76 M_{Sun}$  for the smaller star, consistent with the mass range of $40 M_{Sun}$ - $100 M_{Sun}$  for stars that form massive black holes.\\

\end{document}